\documentclass[fleqn,10pt]{wlscirep}

\usepackage{amsmath}
\usepackage{epstopdf}

\newcommand{\Tr}[1]{\operatorname{Tr} \left\{ #1 \right\}}
\newcommand{\ve}{\varepsilon}

\newcommand{\SO}{{\cal L} } 

\graphicspath{{./Figures/}{"C:/Users/Justin Bergfield/Dropbox/New Josh/"}}

	\title{Signatures of Plexcitonic States in Molecular Electroluminescence}
	
	\author[1,2,*]{Justin P. Bergfield}
	\affil[1]{Department of Physics, Illinois State University, Moulton Hall 311, Normal, IL 61790, USA }
	\affil[2]{Department of Chemistry, Illinois State University, Julian Hall 214, Normal, IL 61790, USA}
	\affil[*]{jpbergf@ilstu.edu}
	\author[3]{Joshua R. Hendrickson}
	\affil[3]{Air Force Research Laboratory, Sensors Directorate, Wright-Patterson Air Force Base, Ohio 45433, USA }

	\begin{abstract}
		We develop a quantum master equation (QME) approach to investigate the electroluminesence (EL) of molecules confined between metallic electrodes and coupled to quantum plasmonic modes.  Within our general state-based framework, we describe electronic tunneling, vibrational damping, environmental dephasing, and the quantum coherent dynamics of coupled quantum electromagnetic field modes.  
		As an example, we calculate the STM-induced spontaneous emission of a tetraphenylporphyrin (TPP) 
		molecule coupled to a nanocavity plasmon.  In the weak molecular exciton-plasmon coupling regime we find excellent agreement with experiments, including above-threshold hot luminescence, an effect not described by previous semiclassical calculations.  In the strong coupling regime, 
		we analyze the spectral features indicative of 
		the formation of plexcitonic states. 

	\end{abstract}
	
\begin{document}
		\flushbottom
		\maketitle
	
\thispagestyle{empty}
	
	\section*{Introduction}
	
	
	The electroluminesence (EL) of individual quantum emitters coupled to metallic electrodes has been investigated extensively 
	since the first scanning tunneling microscope induced luminescence (STML) experiments were performed \cite{gimzewski1988photon,shamai2011spectroscopy}.  Through precise control of an STM probe's position, both the emitter--probe coupling and the resonant frequency of collective motion of the metallic electrons (plasmons) 
	confined in the nanocavity formed between the probe and substrate can be tuned. By adjusting the plasmon frequency, the coupling between particle-hole excitations (excitons) on the quantum emitter and the plasmons can be controlled, leading, for instance, to the observation of plasmon enhanced photon emission 
	\cite{gimzewski1989enhanced,berndt1991inelastic}.  
	%
	This unprecedented control 
	makes STML systems ideal both
	for exploring fundamental aspects of the nonequilibrium electro-optical response of quantum emitters 
	and as a testbed to develop quantum-enhanced device technology, e.g. those related to biological sensing \cite{lee2007exciton}, photovoltaic energy conversion \cite{ferry2010design}, or non-classical light generation \cite{chang2006quantum, tame2013quantum,monroe2002quantum, khitrova2006vacuum,raimond2001manipulating}.  
	
	We focus on molecular emitters in particular 
	since they can be engineered with the specific emission profiles, dipole moments, wavelengths, and symmetries necessary to harness uniquely quantum resources which may be useful in the development of novel opto-electronic devices  \cite{nothaft2012electrically, lounis2005single,aradhya2013single}. 
	When a voltage bias causes the source and drain electrodes' chemical potentials to align with unoccupied and occupied molecular states, respectively,  
	a tunnel current and subsequent molecular exciton are produced.  If the exciton decays radiatively, the resulting EL encodes the specific electronic and vibrational state of the molecule.  Molecular vibrational states have been observed in STML experiments of single porphryin molecules \cite{qiu2003vibrationally,wu2008intramolecular, dong2004vibrationally,dong2010generation,chen2010viewing} and fullerene C$_{60}$ and C$_{70}$ clusters \cite{cavar2005fluorescence,rossel2009plasmon}.
	%
	
	Dong {\em et al.} observed molecular hot-luminescence (HL) from excited vibrational modes in tetraphenylporphyrin (TPP) molecules weakly coupled to metallic electrodes \cite{dong2010generation}.  Their data are a direct observation of the strong dependence of the EL on the resonant frequency of the localized nanocavity plasmons.    In addition, their report of a violation of Kasha's rule, which states that the lowest vibrational transitions should dominate the molecular flourescence, indicates a strong enhancement of the 
	spontaneous emission rate (i.e. a Purcell enhancement \cite{purcell1946spontaneous}) caused by the formation of the nanocavity \cite{le2007surface, chen2015molecular}. 
	Interestingly, above threshold HL (i.e. $eV < \hbar \omega$) was also observed in TPP junctions.\cite{dong2010generation}  In the weak coupling limit, this effect doesn't appear to be described using a classical plasmonic field,\cite{tian2011electroluminescence} although it may be explained when higher-order  electron-plasmon scattering processes are included\cite{kaasbjerg2015theory}.   

	
	In the study of quantum electrodynamics, Purcell enhancement is a signature of the {\em weak coupling} regime between coupled quantum emitters and optical modes.  As the coupling strength is increased 
	there is a transition into the {\em strong coupling} regime, where energy transfers coherently between the emitter and field modes, giving rise to an observable Rabi splitting between the joint emitter-field states.  Systems operating in the strong coupling regime allow for the observation of quantum effects, including single-atom lasing, single photon generation, and all-optical single photon switching \cite{khitrova2006vacuum, mckeever2003experimental, solomon2001single,kimble1998strong,raimond2001manipulating}.   

	Coupled molecular excitons and plasmons form joint states known as plexcitons \cite{savasta2010nanopolaritons,chang2006quantum,tame2013quantum,manjavacas2011quantum}.  Molecular plexcitonic states with Rabi splittings up to several hundred meV have been observed  \cite{bellessa2004strong,dintinger2005strong,hakala2009vacuum,schwartz2011reversible, guebrou2012coherent,fofang2008plexcitonic,vasa2013real,melnikau2016rabi,wang2016role}, motivating a detailed
	investigation into the influence of quantum dynamics, chemical structure, many-body interactions, plexcitonic dynamics, and loss mechanisms on the optoelectronic response of these system.

	
	
	In this article, we develop a state-based quantum master equation (QME) approach to investigate the EL of molecules in both the weak and strong plexcitonic coupling regimes.   
	We first derive an effective multi-state Jaynes-Cummings model for the molecule and quantum plasmon modes, and use the QME framework to describe finite tunneling currents, radiative and non-radiative exciton decay paths, vibrational damping, and finite plasmon lifetimes.   Although similar methods have been used to investigate plasmon-enhanced EL and transport-induced EL in STML systems before \cite{tian2011density,tian2011electroluminescence,zhang2013plasmon,chen2015molecular}, we extend these works to describe the quantum optical regime including a full quantum many-body description of the molecule, plasmon modes, electrodes and their couplings.  As a first application, we simulate the EL of a TPP molecule coupled to a single quantum plasmon mode for several voltages, plexcitonic coupling strengths, and detunings.  

	\begin{figure}[tb]
		\centering
		\includegraphics[width=2.8in]{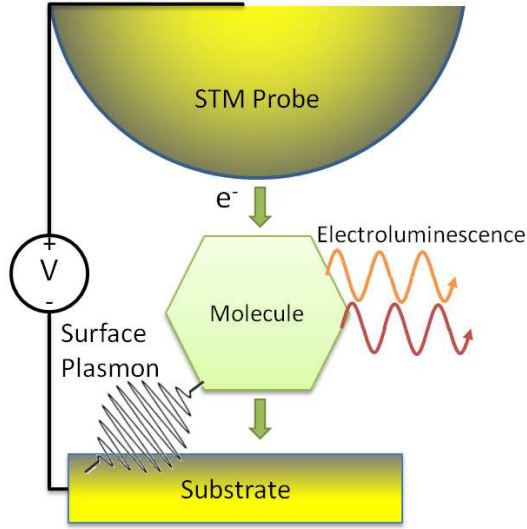}
		\caption{A schematic diagram of an STML experiment including the STM probe, molecule, and substrate.  Molecular excitons generated by a finite tunnel current couple to nanocavity plasmons and may decay radiatively as EL.}
		\label{fig:schematic}
	\end{figure}

	\section*{Theoretical Model}

	%
	%
	%
	
	

	We consider open quantum systems 
	composed of a molecule coupled to electromagnetic field modes, metallic leads (e.g. the substrate and STM probe), and vibrational modes subject to applied voltages and temperature gradients.  
	A schematic of the STM-based experiments we consider is shown in Fig~\ref{fig:schematic}. 
	The Hamiltonian corresponding to this system may be partitioned as 
	\begin{equation}
	H = {H_{\rm mol}} +  {H_{\rm leads}} + H_{\rm vib} + H_{\rm EM} + 
	  H_{\rm tun}  + H_{\rm mol - vib} + {H_{\rm mol - EM}},
	\label{eq:Hamiltonian}
	\end{equation}

	where $H_{\rm mol}$ is the molecular Hamiltonian, and the independent lead, vibrational, and electromagnetic baths are described by 
	\begin{align}
	{H_{\rm leads}} &= \sum_\alpha \sum_{k \in \alpha } {{\varepsilon _{k\sigma }}c_{k\sigma }^\dag {c_{k\sigma }}},  \\ 
	{H_{\rm vib}} &= \sum_\alpha \sum_{l \in \alpha } {{\hbar\Omega_{l }}b_{l\sigma }^\dag {b_{l\sigma }}}, \\
	H_{\rm EM} & = \sum_j \hbar \omega_j a^\dagger_j a_j  
	\end{align}
	respectively, where $c_{k}$ annihilates an electron in lead mode $k$ with dispersion $\varepsilon_{k\sigma}$ and spin $\sigma$, $b_l$ annihilates a vibrational excitation (phonon) in mode $l$ with energy $\hbar \Omega_l$, and $a_j$ annihilates a photon in mode $j$ with energy $\hbar \omega_j$.  
	
	The tunnel coupling, vibrational mode couplings, and electromagnetic field couplings are given by
	\begin{align} 
	H_{\rm tun} & = \sum_\alpha \sum\limits_{k \in \alpha,n\sigma } {{V_{nk}}d_{n\sigma }^\dag {c_{\sigma k}} + H.c.}, \nonumber \\ 
	H_{\rm mol-vib} &= \sum_\alpha \sum\limits_{l \in \alpha,n\sigma } {{W_{nk}}d_{n\sigma }^\dag {b_l} + H.c.},   \nonumber \\
	H_{\rm mol-EM} &=  \int d{\bf r} J({\bf r})\cdot {\bf A}({\bf r}) 
	\end{align}
	respectively, where $V_{nk}$ is the tunneling matrix element, $d_{n\sigma}$ annihilates an electron in molecular state $n$ with spin $\sigma$, and $W_{nk}$ is the vibrational coupling between molecular orbital $n$ and lead mode $k$. 
	The current density $J({\bf r})$ couples to the vector potential ${\bf A}({\bf r})$, which for quantized electromagnetic modes is given by \cite{scully1997quantum}
	\begin{equation}
	{\bf A}_j({\bf r}) = \sqrt{\frac{\hbar}{2\ve_0 \omega_j V}} {\eta}_j({\bf r}) \left(a_j + a_j^\dagger\right)
	\end{equation}
	where ${\bf A} = \sum_j {\bf A}_j$, $\eta({\bf r})$ is the product of the polarization vector and a function describing the spatial profile of the field, $V$ 
	is the effective mode volume, and $\hbar\omega_j$ is the energy of mode $j$.

	\subsection*{Quantum Master Equation}
	In general, the system described by Eq.~\ref{eq:Hamiltonian} cannot be solved exactly.  To proceed, we utilize a state-based quantum master equation (QME) approach, where the quantum dynamics of the joint molecule and plasmon system are treated exactly, while the other macroscopic degrees of freedom are traced over using a coarse-graining procedure.  Within the QME framework, the Liouville equation for the reduced density matrix of the system is given by
	\begin{equation}	
	\dot{\rho} =  - \frac{i}{\hbar}\left[ {{H_0},\rho } \right] +  \left({\cal L}_{\rm tun} + {\cal L}_{\rm damp} + {\cal L}_{\rm deph} \right) \rho, 
	\label{eq:Liouvillian}	
	\end{equation}
	where $\rho$ is the 
	density operator, 
	and the Liouvillian superoperators $\SO_{\rm tun}$, $\SO_{\rm damp}$, and $\SO_{\rm deph}$ describe the nonhermitian evolution of the system due to quantum tunneling, damping of the populations and coherences of states, and pure dephasing, respectively. Once the density matrix is determined, expectation values of observables may be calculated using $\langle {\cal O} \rangle = \Tr{\rho(t) {\cal O}(t)}$.
	%
	%
	
	After coarse graining, our free-system Hamiltonian $H_0$ 
	is composed of three terms: the molecular Hamiltonian, quantum plasmon modes, and their couplings.  The Hamiltonian of each term is given by 
	\begin{align}
	\label{eq:Hfree_start}
	H_{\rm mol} &= \sum\limits_{n,m,\sigma}  {H^{(1)}_{nm}}d_n^\dag {d_m}  + \!\!\! \sum_{ijnm, \atop{\sigma \sigma'}} \!\!\! {{\frac{U_{ijnm}}{2}}d_{i\sigma }^\dag d_{j\sigma '}^\dag {d_{m\sigma '}}{d_{n\sigma }}}  \nonumber \\ & \phantom{abc} + \sum_{l} \hbar \tilde{\Omega}_l \tilde{b}_l^\dagger \tilde{b}_l + \sum_{l,n} \lambda \hbar\tilde{\Omega}_l (\tilde{b}_l + \tilde{b}_l^\dagger) d_n^\dagger d_n,   \\
	H_{\rm plas} &= \sum\limits_j {\hbar \tilde{\omega}_j \tilde{a}_j^\dag {\tilde{a}_j}},    \\
	H_{\rm int} &= \sum_{n,m,j} \hbar g_{nm}^j d^\dagger_n d_m \left(\tilde{a}_j + \tilde{a}_j^\dagger \right),
	\label{eq:Hfree_end}
	\end{align}
	respectively, 
	%
	where $H^{(1)}_{nm}$ is the one-body portion of the molecular Hamiltonian, which may be renormalized by classical electrostatic\cite{barr2012effective} (e.g. image charge) or vibrational effects induced by the electrodes; $U$ is the Coulomb integral, $\tilde{b}_l$ annihilates a (renormalized) phonon in mode $l$ with energy $\hbar \tilde{\Omega}_l$; $\lambda$ is the electron-phonon coupling; and $\tilde{a}_j$ annihilates a plasmon in mode $j$ with energy $\hbar \tilde{\omega}_j$.
	
	For systems we consider, the dipole approximation of the electromagnetic coupling is sufficient. In this approximation
	the plexcitonic coupling parameter is given by \cite{loudon2000quantum}
	\begin{equation}
	\hbar g^j_{nm}  = \sqrt{\frac{\hbar \omega_j}{2\varepsilon_0 V}} \mu_{nm} u_j(x_0)
	\label{eq:g}
	\end{equation}
	where $j$ is the plasmon mode index, $n$ and $m$ are level indices, $V$ is the mode volume, $\mu_{nm} =-e\langle n \left| \vec{r}  \right| m\rangle$ 
	is the transition dipole matrix element, $\omega_j$ is the mode's angular frequency, $\varepsilon_0$ is the permittivity of free space, and $u_j(x_0)$ is the mode function evaluated at the emitter's position $x_0$.

	%
	
	Although we have expressed the molecular Hamiltonian in terms of electron and phonon operators, Eqs.~\ref{eq:Hfree_start}-\ref{eq:Hfree_end} are essentially a multi-state Jaynes-Cummings model where the state energies and matrix elements can be found using a variety of methods (e.g. via exact diagonalization, density functional theory, etc.).

	\subsubsection*{Quantum Transport}
	Electron transport involves the addition and removal of charges which maintain a degree of phase coherence as they traverse a junction. We consider systems 
	in which the molecules and electrodes are deliberately decoupled (e.g. via the growth of insulating layers on the metallic  substrate\cite{qiu2003vibrationally}, or by depositing several molecular monolayers  \cite{dong2004vibrationally}) such that the individual molecule's emission is not quenched by interactions with the metallic electrodes\cite{baffou2008molecular,marinica2013plexciton}. 
	In this 
	regime, the coherence time of electrons on the molecule are short compared to tunneling time, allowing us to neglect the excitation of coherent superposition states and instead describe the transport as a simple kinetic process. 
	\cite{beenakker1991theory,harbola2006quantum}.
	
	Following an expansion of the Liouville equation for the time evolution of the density matrix to second-order in $H_{\rm tun}$,  
	the master equation for tunneling is given by \cite{seldenthuis2010electroluminescence, tian2011density,chen2015molecular}
	\begin{equation}
	{\cal L}_{\rm tun}\rho = \sum_{ij} \left[R_{i \rightarrow j}\sigma_{ji}\rho \sigma_{ij} - R_{j \rightarrow i}\sigma_{jj}\rho \sigma_{jj} \right],
	\end{equation}
	where $R_{i\to j}$ is the charging rate between the $N$-particle state $i$ and the N+1-particle state $j$,  $R_{j\to i}$ is the discharge rate between states $j$ and $i$, and 
	$\sigma_{ij}$ is a matrix in the free system's state space with element $(i,j)=1$ and all other elements equal to $0$.  
	%
	%
	The electronic tunneling rates are given by
	\begin{align}
	R_{i \to j} &= F_{ij} \sum\limits_{\alpha} {\tilde \Gamma _{ij}^\alpha {f_\alpha }\left( {\left[ {E_j - E_i} \right] } \right)},  \\ \nonumber
	{R_{j \to i}} &= F_{ji}\sum\limits_{\alpha} \tilde \Gamma _{ji}^\alpha \left\{ {1 - {f_\alpha }\left( {\left[ {E_j - E_i} \right]} \right)} \right\},   
	\end{align}
	where $F_{ij}$ are the Franck-Condon factors (i.e. the overlap between nuclear wave functions), and $f_\alpha(E)=(1+\exp[(E-\mu_\alpha)/k_{\rm B}T_\alpha])^{-1}$ is the Fermi-Dirac distribution for lead $\alpha$ with temperature $T_\alpha$ and chemical potential $\mu_{\alpha}$.  The chemical potential of the source and drain leads are given by $\mu_S=E_f-e\alpha V_b$ and $\mu_D=E_f-e(1-\alpha)V_b$, respectively, with Fermi energy $E_f$, electron charge magnitude $e$, voltage symmetry $\alpha$, and bias voltage $V_b$. We assume a symmetric potential drop, where $\alpha=0.5$.   
	
	The effective tunnel coupling between electrode $\alpha$ and the molecule is given by \cite{bergfield2009many}
	\begin{equation}
	{\tilde \Gamma}_{ij}^\alpha  = \Tr{ {{\Gamma ^\alpha }\left( {{E_j} - {E_i}} \right)C\left( {i,j} \right)}},
	\label{eq:eff_gamma}
	\end{equation}
	where the bare electron tunneling rate matrix
	$\Gamma ^\alpha_{nm}\left( E \right) = 2\pi/\hbar \sum_{k\sigma  \in \alpha } {{V_{nk}}V_{mk}^*\delta \left( {E - {\varepsilon _{k\sigma }}} \right)}$
	is dressed by the many-body renormalization factors \cite{bergfield2009many}
	\begin{equation}
	{\left[ {C\left( {i,j} \right)} \right]_{n\sigma ,m\sigma '}} = \left\langle j \right|d_{n\sigma }^\dag \left| i \right\rangle \left\langle i \right|{d_{m\sigma '}}\left| j \right\rangle.
	\label{eq:manybody}
	\end{equation}
	As indicated by Eq.~\ref{eq:eff_gamma}, both the relative phase and magnitude of the many-body factors influence the effective tunneling rates and therefore the transport and optical response of systems with multiple states.  In addition to the many-body wave function normalization, where the total resonance width of a molecular state is reduced by a factor of $1/N$ ($N$ is the number of atomic orbitals), strong correlations can also lead to an exponential suppression of these terms.   

	\subsubsection*{Damping and Dephasing}
	
	When a free system interacts with the environment, 
	an initially excited state can decay via a number of irreversible damping processes.  We account for these loss mechanisms with the
	composite Liouvillian operator $\SO_{\rm damp} = \SO_{\rm rad} + \SO_{\rm cav}+\SO_{\rm vib} $, which describes radiative decay processes, 
	the finite lifetime of the nanocavity plasmons, and vibrational relaxation processes, respectively.  

	Assuming Markovian baths, $\SO_{\rm rad}$ 
	can be expressed as a Lindblad master equation \cite{scully1997quantum, seldenthuis2010electroluminescence, tian2011density,tian2011electroluminescence}
	\begin{equation}
	{\cal L}_{\rm rad}\rho  =   \! - \! \sum_{i,j} \!\! \frac{\gamma^{j \to i}_{\rm rad}}{2} \left( \sigma _{ji}\sigma _{ij}\rho  - 2\sigma_{ij}\rho \sigma_{ji} + \rho \sigma_{ji}\sigma_{ij}  \right),
	\end{equation}
	where $\gamma^{j \to i}_{\rm rad}$ is the 
	radiative coupling rate between electronic levels $i$ and $j$.  The finite plasmon lifetime is included via the phenomenological decay rate $\kappa_j$ and 
	master equation
	\begin{equation}
	\SO_{\rm cav} \rho =  - \sum_j {\kappa_j  \over 2 }\left( {{\tilde{a}_j^\dag }\tilde{a}_j\rho  - 2\tilde{a}_j\rho {\tilde{a}_j^\dag } + \rho {\tilde{a}_j^\dag }\tilde{a}_j} \right),
	\end{equation}
	where $\tilde{a}_j$ annihilates a plasmon in mode $j$.

	The intraband vibrational damping may be described by \cite{seldenthuis2010electroluminescence}
	\begin{align}
	{\SO_{\rm vib}}\rho  
	& =  - \gamma_{\rm vib} \sum\limits_i  \left[{ {{\sigma _{ii}}\rho {\sigma _{ii}} - P\left( {\Omega_{\rm vib}^i} \right)\sum\limits_j {{\sigma _{ij}}\rho } {\sigma _{ji}}} } \right],
	\end{align}
	where $\gamma_{\rm vib}$ is the vibrational coupling rate, and $P(\Omega) = e^{-\hbar \Omega / k_B T}/{\cal Z}$, with the partition function ${\cal Z}=\sum_k {e^{-\hbar \Omega^k_{\rm vib} / k_BT}}$.  The states labeled $i$ and $j$ belong to the same electronic manifold.  Although $\gamma_{\rm vib}$ is typically several orders of magnitude larger than the radiative relaxation rate, $\gamma_{\rm rad}$  may be enhanced (e.g. by placing an emitter in a cavity) to exceed the vibrational decay rate, resulting, for instance, in the observation of HL from excited vibrational states 
	\cite{dong2010generation, tian2011electroluminescence,chen2015molecular}. 

	In addition to relaxation, where excitations are transferred from the system to modes of the environment, pure dephasing is also possible, in which the populations are unaffected but their coherence is reduced.
	The Liouvillian term 
	describing 
	dephasing 
	is given by \cite{johansson2005surface} 
	\begin{equation}
	\SO_{\rm deph} = - \gamma_{\rm deph} \sum_{i,j} \left(\sigma_{ij} \rho_{ij} + \sigma_{ji} \rho_{ji} \right)
	\end{equation}
	where $\gamma_{\rm deph}$ is the dephasing rate. 
	%
	%
	
	%

	The relationship between $\kappa$, the total state coupling $\Gamma_{\rm total}$, and the plexcitonic coupling $g$ distinguishes the strong ($g \gg \Gamma_{\rm total}, \kappa$) and weak ($g \ll \Gamma_{\rm total}, \kappa$) coupling regimes.  According to Eq.~\ref{eq:g} the strong coupling regime may be achieved by increasing the transition energy, increasing the molecular dipole moment, or decreasing the effective mode volume accessible to the plasmons  (e.g. by patterning the substrate to exhibit a reduced density of modes).  

	\begin{figure}[tb]
		\centering
		\includegraphics[width=3.5in]{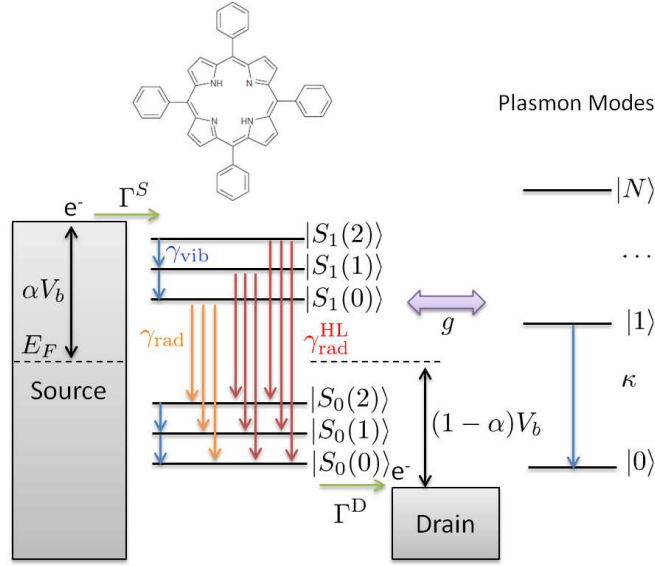}
		\caption{The energy level diagram for the charge neutral manifold of tetraphenylporphyrin (TPP) molecule and coupled quantum plasmon modes.  The relevant (Q-band) electronic states $S_0$ and $S_1$ are shown with their associated vibrational levels.  The observed optical gap of TPP is 1.89eV, with a vibrational state spacing of 0.16eV.\cite{dong2010generation}  Based on comparison with experiment, the source and drain tunnel couplings are set to $\hbar \Gamma^{\rm S}=\hbar \Gamma^{\rm D}$=16.4$\mu$eV, while the vibrational damping $\gamma_{\rm vib}$, and radiative decay  $\gamma_r$, are consistent with a vibrational lifetime of 2ps and a radiative lifetime of 2ns.\cite{dong2010generation} 
			We set the plasmonic decay to $\hbar \kappa=10meV$ and plexcitonic coupling $g$ is taken as an adjustable parameter.}
		\label{fig:energy_level}
	\end{figure}

	
	\begin{figure*}[htb]
		\centering
		\includegraphics[width=.44\linewidth]{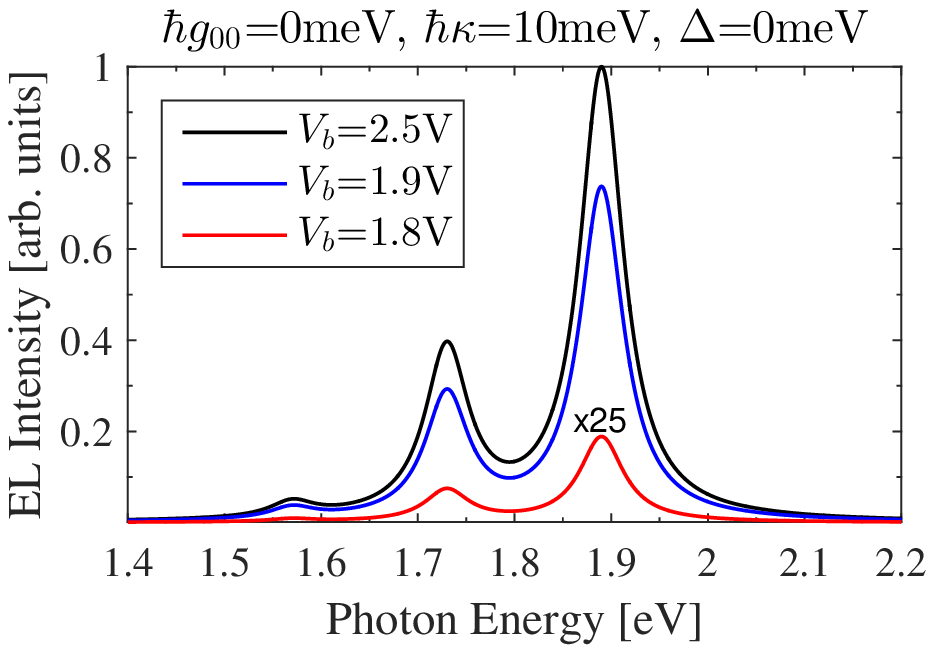} 
\includegraphics[width=.44\linewidth]{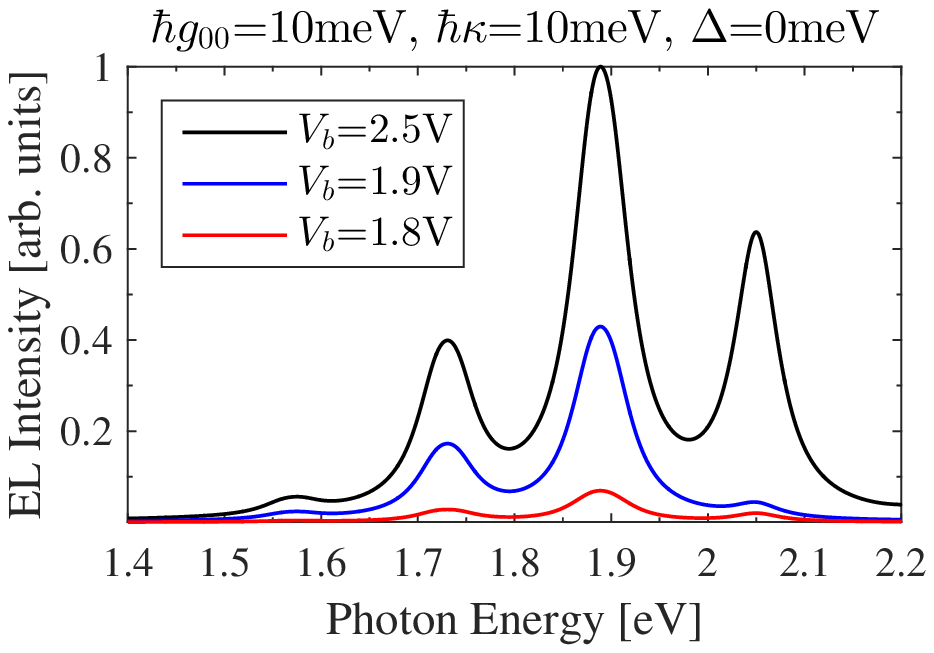} 
\includegraphics[width=.44\linewidth]{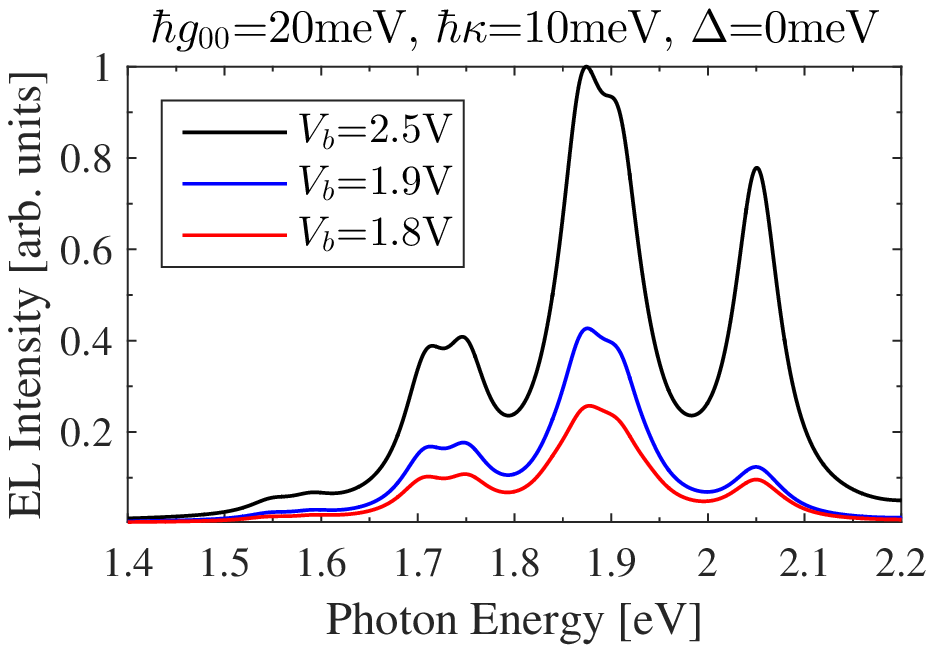} 
\includegraphics[width=.44\linewidth]{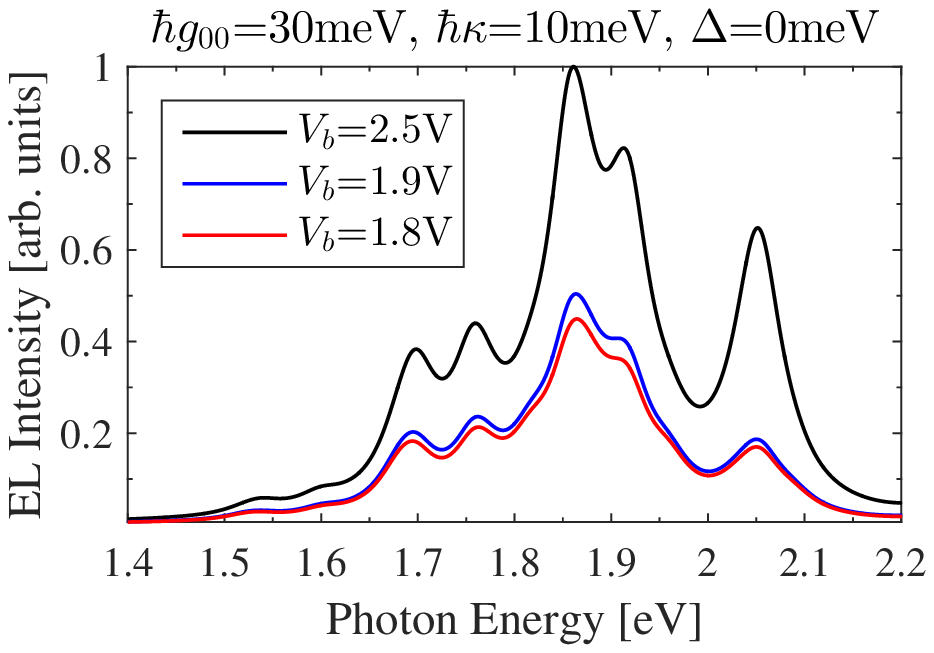} 
\vspace{-0.5cm}
		\caption{The calculated STML of a TPP molecule coupled to a single plasmon mode is shown as a function of photon energy for four plexcitonic coupling strengths $\hbar g_{00}$=0meV, 10meV, 20meV, and 30meV at three bias voltages $V_b$=1.8V, 1.9V, and 2.5V. In the decoupled cavity limit ($\hbar g_{00}=0$meV), the molecular vibrational spectrum is recovered.  As $g_{11}$ is increased plexcitonic states form giving rise to the characteristic split peaks in the EL.  Although all radiative transitions have non-zero coupling to the plasmonic field, the resonant $Q(0,0)$ transition dominates and the plexcitonic splitting is $\sim2\hbar g_{00}$.  The coherent mixing of multiple states gives rise to the asymmetric peak structure and the above threshold hot HL observed when $V_b$=1.8V; even for junctions operating in the weak coupling regime.  For this system, the strong coupling boundary is $\hbar g_{00} > \sim$15meV.  Calculations are for junctions operating at $T$=80K, to be consistent with measured STML  spectra.\cite{dong2010generation}}
		\label{fig:resonant_TPP_vs_g}
	\end{figure*}

%
	\section*{Results and Discussion}


	As a first test of our theory, we investigate the STML of a tetraphenylporphyrin (TPP) molecule coupled to a single quantum plasmon mode.  In general, the spontaneous emission spectrum may be found from the density matrix though the use of the quantum regression theorem \cite{gardiner2004quantum,laussy2009luminescence}.  Since our model for the TPP molecule is constructed {\em ad hoc} from experimental data, we consider it to be a sum of Lorentzians given by
	\cite{tian2011density,tian2011electroluminescence,chen2015molecular}
	\begin{equation}
	I(\omega) =  \frac{1}{2\pi} \sum_{i,j} \frac{\gamma_{\rm rad}^{j\rightarrow i} \rho^{jj}_{ss}}{(\omega - \omega_{ji})^2 +\tau_{ji}^2}
	\end{equation}
	where $\gamma_{\rm rad}^{j\rightarrow i}$ is the radiative decay rate between states $j$ and $i$,  $\rho_{ss}$ is the steady-state solution of Eq.~\ref{eq:Liouvillian}, and $\tau_{ji}$ is the full width at half maximum (FWHM) of the EL.  We assume that the plexcitonic EL linewidths are well approximated by the zero-detuned single atom vacuum Rabi splitting linewidths $\tau_{ji}=(2\tau_{ji}^0 + \kappa) / 2$, where $\tau_{ji}^0$=0.05eV is the FWHM extracted from experiments operating in the weak coupling limit \cite{dong2010generation,tian2011density}.
	

	A schematic of the TPP molecule and the energy level energy diagram for the TPP system are shown in Fig.~\ref{fig:energy_level}.  The $Q$-band 
	energy gap of TPP (i.e. the $Q(0,0)$ transition, $S_1(0)\rightarrow S_0(0)$) is 1.89eV, with a vibrational level spacing of $0.16$eV. \cite{dong2010generation, gouterman1978optical}  We consider the TPP molecule coupled to a single quantum plasmon mode with energy $\hbar \tilde{\omega}$ = 1.89eV -  $\Delta$, where $\Delta$ is the detuning parameter. The source and drain tunnel coupling rates ($\Gamma^{\rm S}=\Gamma^{\rm D}=0.4\times10^{10}s^{-1}$), vibrational decay rates ($\gamma_{\rm vib}=0.5\times10^{12}s^{-1}$), and radiative decay rates ($\gamma^{j \rightarrow i}_{\rm rad} = F_{ji} / 2ns$)
	are established by comparison with experiment. \cite{tian2011density,tian2011electroluminescence,dong2010generation}  
	The transition rates between vibrational levels and the transition dipole moments are scaled by the appropriate Franck-Condon factors $F_{ij}$, which are found using the harmonic approximation \cite{keil1965shapes} with a Huang-Rhys parameter $S=0.61$ (See Supporting Information).\cite{tian2011electroluminescence}  We assume that the radiative lifetime of all states are equal (i.e. $\gamma_{\rm rad}^{\rm HL}=\gamma_{\rm rad}$), and calculations were performed using a modified version of QuTIP. \cite{johansson2012qutip}

	
	%


	

	\begin{figure*}[htb]
		\centering
		\includegraphics[width=.48\linewidth]{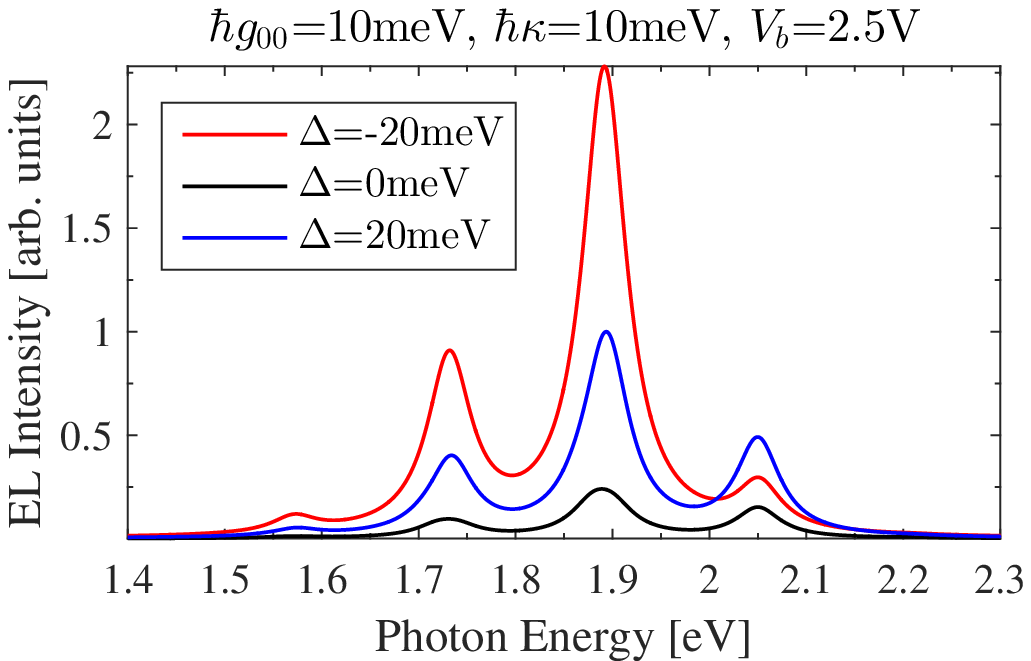}
		\includegraphics[width=.48\linewidth]{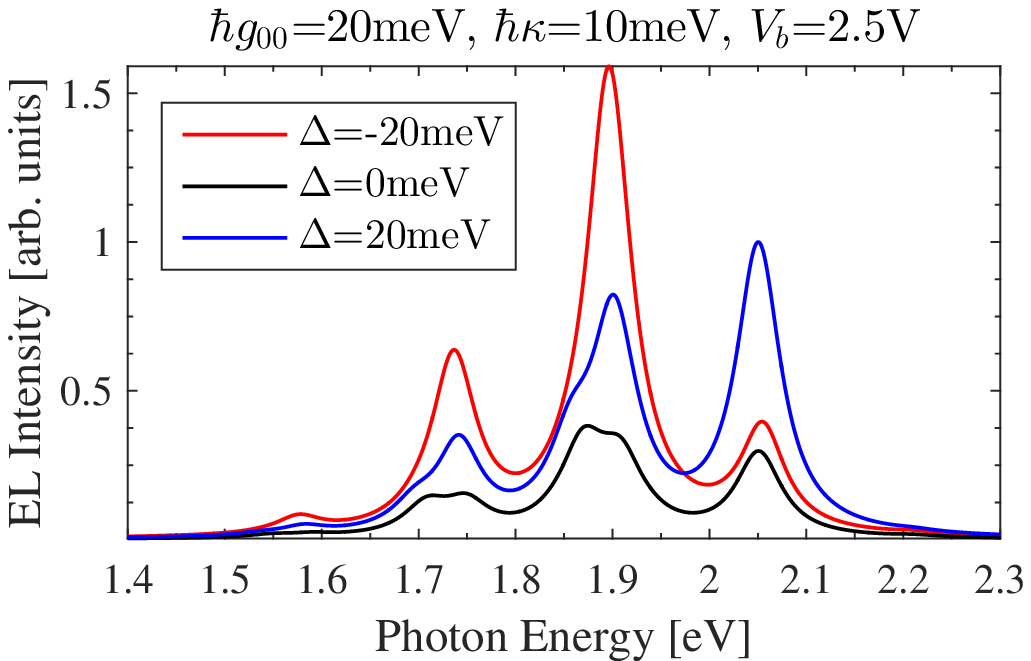}
		\caption{The calculated STML of a TPP molecule coupled to a single plasmon in the weak (left panel) and strong (right panel) coupling regimes for several detuning values.  The EL is enhanced by detuning the plasmon and molecular excitation resonances since the plasmonic nanocavity lifetime is limited by $\kappa$.  In the strong coupling regime ($|g_{00}|/\kappa > 0.25$), detuning can be used to favor the upper or lower plexcitonic state.  Simulations are for junctions operating at $T$=80K. 
		}
		\label{fig:EL_detune}
	\end{figure*}

	
	
	
	
	For a resonant plasmon, the plasmonic decay rate $\kappa$ may be expressed in terms of the mode energy $\omega$ and quality factor $Q$ as $\kappa = \omega / Q$.  Using reports for other metal-insulator-metal nanostructures, \cite{yang2012ultrasmall} we find that values of $Q \approx 100$ are reasonable.  Given that a plasmonic mode's lifetime (and the cavity quality factor) can vary significantly for different junction designs, probe positions, and substrate materials, we consider $Q$=189 in our calculations, such that $\hbar\kappa$=10meV.
	
	

	The calculated STML of a TPP junction for four plexcitonic couplings and three bias voltages are shown as a function of photon energy in  Fig.~\ref{fig:resonant_TPP_vs_g}.  
	In the decoupled cavity limit (top left panel, $\hbar g_{00}$=0meV), the experimentally observed spontaneous emission peak structure \cite{dong2010generation} is recovered, where the three peaks corresponding to the $Q(0,0)$, $Q(0,1)$, and $Q(0,2)$ transitions of the TPP molecule.  As $g_{00}$ is increased, the molecular exciton and cavity plasmon states mix, forming plexcitonic states separated in energy by $\sim 2 \hbar g_{ij}$.  In the weak coupling regime the radiative transition rate is enhanced via the Purcell effect, making HL possible, while in the strong coupling regime energy exchanges coherently between plasmonic and molecular states resulting in a characteristic split peak structure of the EL.  The boundary between weak and strong coupling regimes is defined by $|g_{00}|/|\kappa_t| = 0.25$, where for the TPP junction $\kappa_t \approx \kappa + \tau_{ij}^0$ = 60meV and therefore $\hbar g_{00} = \sim$15meV.    
	
	The STML of a TPP junction operating in the weak coupling regime is shown in the top right panel of Fig.~\ref{fig:resonant_TPP_vs_g}, where a HL peak at 2.05eV is visible for all bias voltages.  Although the peak was observed experimentally at 1.8V, where $eV_b$ is less than the excitation energy of the molecule,\cite{dong2010generation} it was not seen at this voltage in previous calculations using classical plasmonic fields.\cite{tian2011electroluminescence}  This implies that Purcell enhancement alone can't explain the measured HL.  
	
	The strong coupling STML of the TPP junction 
	is shown in the lower two panels of Fig.~\ref{fig:resonant_TPP_vs_g}, where the $Q(0,0)$ and $Q(0,1)$ peaks have split into two peaks.  The asymmetry of these plexcitonic peaks stems from the influence of multiple detuned resonances, where the plexcitonic coupling between levels is reduced by the appropriate Franck-Condon factor. The $Q(0,2)$ and HL peaks are also split but can't be identified with $\kappa=10meV$ due the reduced effective couplings.

	Our simulations show that HL is suppressed when the off-diagonal coupling terms $g_{i \neq j}$ are reduced or, as expected,  
	 when the vibrational relaxation rate is increased.  This suggests that the HL peaks are a consequence of the (weak) coherent dynamics between off-resonant states.  Although the above-threshold emission has been explained in terms of higher-order many-body processes \cite{tobiska2006quantum,schneider2010optical, kaasbjerg2015theory}, our calculations support an additional physical explanation in which the tunnel current pumps energy into cavity modes via the nascent plexcitonic states.  As shown in the lower panels of Fig.~\ref{fig:resonant_TPP_vs_g}, when the plexcitonic coupling is increased the detuning between states is reduced and, consistent with our argument, the above-threshold emission is enhanced.
	Finally, we consider the influence of the detuning between molecular transitions and  cavity plasmon resonance energies on the EL.  Physically, detuning can be controlled by adjusting the STM probe's height above the substrate.  As evidenced by the peak at 2.05eV shown in the left panel of Fig.~\ref{fig:EL_detune}, Purcell enhancement in the weak coupling regime results in HL for all detunings.  When the cavity is blue or red detuned relative to the $Q(0,0)$ transition of TPP, the STML spectral weight is shifted towards higher or lower energy peaks, respectively.  Since the TPP junction supports a finite current, 
	energy is constantly (albeit weakly) pumped into off resonant cavity plasmon modes, giving the observed shift to the spectrum.  The EL is {\em increased} by detuning since detuning reduces the molecule's effective coupling to non-radiative plasmonic decay paths. 
	
	In the strong coupling regime, shown in the right panel of the same figure, the blue and red detuned plasmon modes again shift the STML spectral weight up or down in energy, respectively.  However, in this regime the strongest peaks are split into distinguishable plexcitonic resonances.  In addition to the characteristic split peak EL, plexcitonic states and the onset of the strong coupling regime can also be identified via this distinct spectral weight shift with detuning.

	\section*{Conclusions}

	We develop a QME approach to investigate the STML of molecules coupled to quantized electromagnetic modes. Within our method we include the effects of electronic tunneling, vibrational damping, and environmental dephasing, and can describe both weak and strong plexcitonic coupling regimes. Our approach extends existing methods and includes a full quantum description of the coherent state dynamics.  Our method is valid in both single-particle and many-body representations, allowing future studies to balance computational effort with chemical accuracy.
	
	Motivated by the observation of HL in the STML of TPP,\cite{dong2010generation} and the argument that it was a consequence of STM-induced Purcell enhancement \cite{chen2015molecular,tian2011electroluminescence,le2007surface}, we calculated the EL of a TPP molecule coupled to a single quantum plasmon mode.  In the weak coupling regime, we recover the experimentally observed spectra, including the above-threshold HL.  Using a fully quantum plasmon theory, we conclude that the low-bias HL peak may be a consequence of the weakly coherent energy exchange dynamics.  Finally, we identify several signatures of the formation of plexcitonic states: a split peak structure of the EL, and the shifted spectral weight as the plasmon resonance is tuned. 
%

	Although the strong coupling regime has not yet been observed in STML systems, molecular systems with coupling strengths of hundreds of meV have been fabricated\cite{bellessa2004strong,dintinger2005strong,hakala2009vacuum,schwartz2011reversible, guebrou2012coherent,fofang2008plexcitonic,vasa2013real,melnikau2016rabi,wang2016role}.  For the TPP system investigated here, we find  that it is physically plausible to achieve the strong coupling regime if the nanocavity losses are reduced slightly, e.g. via careful material selection or patterning of the substrate to reduce the effective plasmon mode volume (See Supporting Information).  If realized, the ability to measure the spatial distribution of the electro-optical response of molecules operating in the strong coupling regime would be invaluable in the development of myriad quantum information applications, and would herald a new phase in the study of QED and molecular dynamics. 

\section*{Acknowledgments}
We acknowledge support from the Air Force Office of Scientific Research (Program Manager Dr. Gernot Pomrenke) under contract number 15RYCOR159.

\section*{Author contributions statement}
J.P.B and J.R.H conceived the idea and performed the theoretical analysis together.  J.P.B. developed the quantum transport theory and code.  Both J.P.B. and J.R.H. wrote the manuscript.

\section*{Additional information}
\textbf{Competing financial interests}: The authors declare that they have no competing interests. \\
\textbf{Supplementary Information}: Details of the Franck-Condon calculations, estimates of the Rabi splitting in TPP, and room temperature STML of TPP can be found in the supporting information



\bibliography{refs}

\end{document}